\documentclass[superscriptaddress,twocolumn,showpacs,preprintnumbers,prb]{revtex4}
\usepackage{graphicx}
\usepackage{times}
\usepackage{epsfig}

\usepackage{color}

\begin{document}

\def\salto{\vskip 1cm}
\def\lgr{\langle\langle}
\def\rgr{\rangle\rangle}

\title{The spatial range of the Kondo effect: a numerical analysis}
%
\author{C.~A. B\"usser}
\email[Corresponding author: ]{carlos.busser@gmail.com}
\affiliation{Department of Physics, Oakland University, Rochester,
MI 48309, USA}
\author{G.~B. Martins}
\affiliation{Department of Physics, Oakland University, Rochester,
MI 48309, USA}
\author{L. Costa Ribeiro}
\affiliation{Departamento de F\'{\i}sica, Pontif\'{\i}cia
Universidade Cat\'olica do Rio de Janeiro, 22453-900, Brazil}
\author{E. Vernek}
\affiliation{Instituto de F\'isica - Universidade Federal de Uberl\^andia - Uberl\^andia, 
MG 38400-902 - Brazil}
\author{E.~V. Anda}
\affiliation{Departamento de F\'{\i}sica, Pontif\'{\i}cia
Universidade Cat\'olica do Rio de Janeiro, 22453-900, Brazil}
\author{E. Dagotto}
\affiliation{Materials  Science and Technology Division, Oak Ridge
 National Laboratory,
 Oak Ridge, Tennessee 37831, USA and\\
 Department of Physics and Astronomy, University of Tennessee,
 Knoxville, Tennessee 37996, USA}

\begin{abstract}
{The spatial length of the Kondo screening is still a controversial 
issue related to Kondo physics. While renormalization group 
and Bethe Anzats solutions have provided  detailed 
information about the thermodynamics of magnetic impurities, they are 
insufficient to study the effect on the surrounding electrons, i.e., the 
spatial range of the correlations created by the Kondo effect between the localized 
magnetic moment and the conduction electrons. 
The objective of this work is to present a quantitative way of measuring 
the extension of these correlations by studying their effect directly on the local 
density of states (LDOS) at arbitrary distances from the impurity.
The numerical techniques used, the Embedded Cluster Approximation, 
the Finite U Slave Bosons, and Numerical Renormalization Group, calculate the Green functions
in real space. With this information, one can calculate how the local 
density of states away from the impurity is modified by its presence, below and above 
the Kondo temperature, and 
then estimate the range of the disturbances in the non-interacting Fermi sea due to the Kondo effect, 
and how it changes with the Kondo temperature $T_{\rm K}$. 
The results obtained agree with results obtained through spin-spin correlations, 
showing that the LDOS captures the phenomenology of the Kondo cloud as well. 
To the best of our knowledge, it is the first time that the LDOS is used to estimate the extension of the Kondo cloud. 
}
\end{abstract}

\pacs{73.23.Hk, 72.15.Qm, 73.63.Kv}
\maketitle

\section{Introduction}
\label{sec:introduction}

The physics of isolated impurities inside a Fermi sea has received
considerable attention since it was experimentally shown that nano-systems
composed by quantum dots possess Kondo phenomena, very clearly
reflected in its transport properties.\cite{david-nature} One
signature of this effect is a narrow resonance, at the Fermi energy, in the
local density of states (LDOS) of the impurity, with a width of the order 
of a characteristic energy, the so-called Kondo temperature, $T_{\rm K}$. 
The transport properties of a nanoscopic
structure in this regime are substantially affected by the Kondo
resonance, as it creates an extra channel at the Fermi level through
which the electrons can propagate. The energy $k_B$$T_{\rm
K}$ is also associated with
an antiferromagnetic correlation between the impurity and the
conduction electron spins in its neighborhood, favoring the emergence
of a singlet ground state. These spins, localized in the impurity's 
vicinity, constitute a screening cloud 
of the localized impurity spin, known as the Kondo cloud. While most of
the physics involved in this 
important effect is by now well established, the nature, structure, and extension of 
the Kondo cloud, and even its existence, is still, to some extent, 
controversial.\cite{Affleck5,sorensen-affleck} 
Theoretically, it is thought to be a crucial ingredient in helping to understand, 
for instance, the interaction between two nearby impurities, when one of them is 
sitting within the region of influence of the Kondo cloud of the other. 
From the experimental point of view, although it is thought that the extension of this 
characteristic cloud can reach very large values,\cite{Affleck5} the properties of 
a system of impurities in metals seem to depend linear on the impurity concentration. 
This seems to indicate that the impurities do not see each other, although, based on 
the expected Kondo cloud extension, they should. Moreover, there has not been any 
clear experimental evidence of it existence, with the exception of the electronic conductance 
measurements in quantum corrals.\cite{manoharan} For instance, in an ellipsoidal quantum corral, 
a Kondo peak produced by a magnetic atom located at one focus of the ellipse has 
an experimentally detectable spectral response in the other focus. This indicates 
a very peculiarly structured Kondo cloud, which, through the use of an 
Scanning Tunneling Microscope (STM), can be experimentally analyzed.

The Kondo cloud length can be estimated by considering that the mean-life of the Kondo
quasi-particles are related to the time-scale $\tau_{\rm K} \approx
\hbar/k_B T_{\rm K}$. Assuming that these quasi-particles propagate with the
Fermi velocity $v_F$, then the Kondo screening length can be
related to the quantity\cite{Affleck5}
\begin{equation}
R_{\rm K} \approx \frac{\hbar v_F}{k_B T_{\rm K}}. \label{R_K}
\end{equation}
Obviously, since all electrons whose energies fall within the Kondo peak 
will participate in the formation of the Kondo cloud, the quantity $v_F$ is not well
defined. Moreover, one may expect that the quasi-particles do not
propagate with the bulk $v_F$, but with a renormalized $v^*$, given by the
presence of the impurity. From Heavy Fermion theory, we can estimate
$v^*=k_F/m^*$, where $m^*$ is the effective mass of the
quasi-particle.\cite{note1} Therefore, Eq.~\ref{R_K} should provide inaccurate
results for the screening length $R_{\rm K}$. However, it should be
expected to give the correct dependence with $T_{\rm K}$ and some
plausible order of magnitude for its length.

From the theoretical point of view, this problem has been analyzed
using different approaches.\cite{Gubernatis,Affleck0,Bergmann1,Gazza1} 
The study of spin properties, through the
local susceptibility or the spatial spin correlation function, has 
given significant contributions to the understanding of this
phenomenon.\cite{Gubernatis,LBorda} More closely related to our approach, the
analysis of the conductance of a quantum dot embedded in a finite
wire,\cite{Affleck5} or the persistent currents in a finite 
ring,\cite{sorensen-affleck} using
renormalization arguments or Density Matrix Renormalization Group calculations, respectively, were 
proposed as a way of determining the Kondo cloud, as well. 

More recently, a variational approach was proposed to 
study the propagation, from the impurity, of the local hole-density.\cite{Simonin}
In this work, it was possible to show that, in two and three dimensions, 
the extension of the Kondo cloud is of 
the order of a few Fermi wave-lengths only, due to angular
dispersion effects, such that $R_{\rm K}$ does not play a significant role in the
physics of a system of impurities in either of these dimensions. This seems to
explain the situation from the experimental point of view (as mentioned above), 
and thus the irrelevance of the Kondo cloud in most of the real systems studied.
However, for one-dimensional systems,\cite{Simonin} the
impurity-impurity
interaction should be determined by the Kondo cloud length $R_{\rm K}$. This will have 
important consequences to the conductance properties, and
therefore will have implications to the design of quantum 
dot integrated nanoscopic systems.

A. Holzner {\it et al.}, using DMRG, have calculated the spin-spin correlations 
involved in the formation of the Kondo cloud in a one-dimensional system and found 
the dependence of the range of the Kondo cloud with the Kondo temeperature to 
agree with Eq.~\ref{R_K}.\cite{fabian}

Motivated by this situation, we study the Kondo cloud in a one-dimensional 
system, focusing our attention on its electronic properties. 
The study of the propagation of the 
Kondo resonance, located at the vicinity of the Fermi energy,  will
shine light, for instance, into the transport properties of a 
quantum dot connected to leads where the distance from the dot
to an STM tip is changed in a controlled and continuous way.\cite{crommie} 
This will be experimentally similar to the transport properties studies
of a system formed by a magnetic atom located in one focus of
an elliptical quantum corral, as mentioned above, and can be
experimentally implemented for a quantum dot connected into an
infinite wire.

In the present work, we discuss the spatial behavior of the Kondo cloud by alternative
means in an infinite one-dimensional system. Indeed, to estimate the
cloud range, three different numerical techniques are used to track
the effects of the impurity over the LDOS
far away from the impurity. These effects are calculated above and
below the Kondo temperature $T_{\rm K}$, and their difference is
used as a {\it finger print} of the extension of the Kondo cloud. The
calculations are carried out using the Embedded  Cluster
Approximation (ECA),\cite{ECA} the Finite-U Slave Bosons Mean-Field
approximation (FUSBMF),\cite{kotliar} and the Numerical
Renormalization Group (NRG) method.\cite{Wilson75}

The paper is organized as follows: In the next section
(Sec.~\ref{sec:system}), we present the model used and the methods to
solve it. In Sec.~\ref{sec:LDOS}, we briefly describe the behavior of
the LDOS, in real space, within the metal lead and define the
function used to estimate $R_{\rm K}$. In Sec.~\ref{sec:results}, the
numerical results calculated using ECA, FUSBMF and NRG are discussed
and compared. Finally, in the last Section, we present our
conclusions.

\section{system and numerical methods}
\label{sec:system}

\begin{figure}
\epsfxsize=8.5cm \centerline{\epsfbox{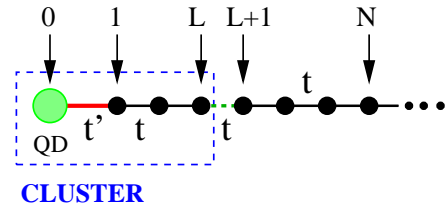}}
\caption{(Color online) Schematics of the system used to determine the spatial extension of the
Kondo cloud. The exactly diagonalized (ED) cluster (with a variable number of sites 
L+1, where the QD is numbered as site zero) is indicated by the dashed box. 
Note that the (integer) index $0\leq N < \infty$ runs over all 
sites {\it inside and outside} the cluster.
By applying the numerical methods employed in this work, the dressed (interacting) 
Green function for all sites in the cluster is obtained (note that for FUSBMF and NRG, the 
cluster can be considered as just the impurity -- see end of Sec.~\ref{sec:EOM}B for details). 
A simple Equation of Motion 
procedure allows the calculation of the GF in any site of the semi-chain 
{\it outside} the cluster. With that, the effect of the Kondo screening effect over the non-interacting 
Fermi sea, the so-called Kondo cloud, can be probed at an arbitrary distance from the impurity.
}
\label{fig1}
\end{figure}

In this work, we analyze a system composed by one Anderson impurity 
(representing, for example, a quantum dot (QD) or an add-atom in a metal surface) coupled 
by a matrix element $t'$ to a band (modeled by a semi-infinite 
non-interacting chain -- from now on referred to as a semi-chain). This system is 
shown schematically in Fig.~\ref{fig1}. 
This figure presents the system using the 
terminology appropriate for ECA, where a finite cluster has to be defined. However, 
as described in more detail below, 
we will show that, to calculate the Green functions (GF) {\it outside} this cluster, when 
doing FUSBMF or NRG, the same terminology can be used, although there is no equivalent 
cluster definition in FUSBMF or NRG. 

As shown in Fig.~\ref{fig1}, $N$ is an (integer) index that numbers the sites from zero to 
infinity, being the impurity (green circle), or QD, located at site $N=0$. The letter $L$ is 
not an index and its meaning, related to ECA, but extended to the other methods, is explained 
below in Sec.\ref{sec:ECA}.

The total Hamiltonian reads,
\begin{eqnarray}
H_{T} &=& H_{\mbox{\small imp}}+ H_{\mbox{\small band}} + H_{\mbox{\small hybrid}},
\end{eqnarray}
with 
\begin{eqnarray}
H_{\mbox{\small imp}} &=& V_g \sum_{\sigma} n_{d\sigma} + U/2 \sum_{\sigma} n_{d\sigma} n_{d\bar{\sigma}} \\
H_{\mbox{\small band}} &=&  t\sum_{N=1\sigma}^\infty (c_{N\sigma}^{\dagger}
c_{N+1\sigma} + c_{N+1\sigma}^{\dagger} c_{N\sigma}) \\
H_{\mbox{\small hybrid}} &=& t' \sum_\sigma (c_{0\sigma}^{\dagger} c_{1\sigma} + c_{1\sigma}^{\dagger}
c_{0\sigma}
),
\end{eqnarray}
where $c_{0\sigma}^\dagger$ creates an electron at the impurity, 
$c_{N\sigma}^\dagger$ creates an electron at the site $N$ of the lead, 
and $n_{d\sigma}= c_{0\sigma}^\dagger c_{0\sigma}$ is the number
operator at the impurity. The first two terms of $H_T$ represent
the Hamiltonian of the impurity and the non-interacting band, respectively, 
and the last term is the hybridization between them.
An important quantity for this system is the broadening of the impurity 
level $\Gamma=2\pi t'^2 \rho_{\mbox{\small lead}}(E_F)$, where
$\rho_{\mbox{\small lead}}(E_F)$ is the LDOS of the first site in the
semi-chain at the Fermi energy $E_F$. Note 

In this paper, we will concentrate our attention on the electron-hole symmetric point 
($V_g=-U/2$), although the results can be generalized to an arbitrary value of 
gate potential. As mentioned in the introduction, we want to estimate 
the extension of the Kondo cloud, and its dependence on $U/\Gamma$, 
by analyzing the LDOS calculated through 
the local GF. To calculate the GF, we will use the ECA, the FUSBMF, 
and the NRG methods, which are briefly described next. 

At the end of this section we will describe how the LDOS in a site $N$ far away 
from the impurity is calculated using the Equation of Motion (EOM) method.
Note that the EOM method described in Sec.~\ref{sec:EOM} 
does not depend on the method used to calculate $G_{dd}^\sigma$; 
any of the three methods described below provide essentially the same kind of input 
for the EOM procedure.

\subsection{Numerical methods}
\label{sec:methods}
\subsubsection{Embedded Cluster Approximation method}
\label{sec:ECA}
The ECA method has been developed to treat localized impurity systems
consisting of a many-body interacting region weakly coupled to
non-interacting conduction bands. The approach is based on the idea
that the many-body effects of the impurity are local in character
(the Kondo cloud, for instance). With this in mind, we proceed in
three steps: first, out of the complete system (the impurity plus a 
non-interacting band, described by a tight-binding Hamiltonian),
one isolates a cluster consisting of the impurity plus their $L$
nearest neighboring sites in the tight-binding semi-chain. 
This cluster, with a variable size $L+1$, as it includes the impurity, 
is shown in Fig.~\ref{fig1} by dashed lines. The first site 
outside the cluster is labeled $N=L+1$ (remember that $N$ is an index, 
as opposed to a number of sites).
Most of the many-body effects are expected to be confined to this cluster. 

The second step of the method 
consists in exactly diagonalizing the cluster, using, for example, 
the Lanczos method,\cite{Elbio} and calculate all the GFs. Finally, 
in the third step, the cluster is embedded into the rest 
of the tight-binding semi-chain using a Dyson equation.\cite{ECA}

Being ${g_{i,j}}$ a cluster-GF that propagates an electron from
site $i$ to $j$ and ${g_{L+1}}$ the GF of the first site out of 
the cluster ($N=L+1$), the Dyson equation to calculate a dressed
(by the presence of the semi-chain), GF for sites inside 
the cluster can be written as,
\begin{eqnarray}
G_{i,i} &=& g_{i,i} ~+~ g_{i,L} ~t~ G_{L+1,i} \label{Dyson1}
, \\
G_{L+1,i} &=& g_{L+1} ~t~ G_{L,i}~;
\end{eqnarray}
Note that the hopping parameter $t$ in Eq.~\ref{Dyson1} corresponds to the
broken link shown by a dashed line in Fig.~\ref{fig1}. This matrix element 
has the same value as all the other hopping parameters within the chain.
Properties like conductance through the impurity and its LDOS, for example,
can be obtained solving this set of equations (for more details, 
see Ref.~\onlinecite{ECA}).

\subsubsection{Finite-U slave bosons mean-field approximation}
The slave boson mean field is a method proposed originally to treat
the problem when the Coulomb repulsion $U$ is the larger quantity.
The double occupancy is excluded from the Hilbert space
with the help of projectors-bosons operators. After taking a mean field
in the boson operators, the many-body Hamiltonian is mapped into an
effective one-body Hamiltonian that can be solved exactly.\cite{newns} 

The FUSBMF approach is an extension of the usual slave boson mean field
in order to treat problems with finite $U$.\cite{kotliar} 
The first step is to enlarge the Hilbert space,
by introducing a set of slave boson operators ${\hat e}$, $\hat
p_{\sigma}$ and $\hat d$, and replacing the creation ($d^\dagger_{\sigma}$) 
and annihilation ($ d_{i\sigma}$) operators in
the Hamiltonian by $d^\dagger_{\sigma}\hat z^\dagger_{\sigma}$ and
$\hat z_{\sigma}d_{i\sigma}$, respectively. Following Kotliar and
Rukenstein,\cite{kotliar} the operator $z$ takes the form\cite{kotliar}
\begin{eqnarray}
\hat z_{\sigma}&=&[1-\hat d^\dagger\hat d-\hat p^\dagger_{\sigma} \hat p_{\sigma}]^{1/2}
[\hat e^\dagger\hat p_{\sigma}+\hat p^\dagger_{\bar\sigma}\hat d]
\nonumber \\ &&
\times[
1-\hat e^\dagger\hat e-\hat p^\dagger_{\bar\sigma}\hat p_{\bar\sigma}]^{1/2}.
\end{eqnarray}
Notice that the bosonic operators $\hat d$ and $\hat d^\dagger$ do not carry an spin index.
The enlarged Hilbert space is then restricted to the physically meaningful 
subspace by imposing the constraints
\begin{equation}
\hat P=\hat e^\dagger \hat e+\sum_\sigma
\hat p^\dagger_{\sigma}\hat p_{\sigma}+\hat d^\dagger\hat d-1=0
\label{cP}
\end{equation}
 and
\begin{equation}
\hat Q_{\sigma}=n_{d\sigma}-\hat p^\dagger_{\sigma}\hat p_{\sigma}
-\hat d^\dagger \hat d=0.
\label{cQ}
\end{equation}
Both constraints are included into the Hamiltonian through Lagrange 
multipliers $\lambda^{(1)}$ and $\lambda^{(2)}_{\sigma}$.  
The constraint described by Eq.~\ref{cP} force the dots to have empty, 
single, or double occupancy only, while the constraint of Eq.~\ref{cQ} 
relates the boson with the fermion occupancies. 
In the mean-field approximation, we replace the boson operators $\hat e$, $\hat
p_{\sigma}$ and $\hat d$ (and their Hermitian
conjugates) by their thermodynamical expectation values $e\equiv\langle\hat e\rangle=
 \langle\hat e^\dagger\rangle$,
 $p_{\sigma}\equiv\langle\hat p_{\sigma}\rangle=
 \langle\hat p^\dagger_{\sigma}\rangle$
and  $d\equiv\langle\hat d\rangle=\langle\hat
d^\dagger\rangle$. These expectation values, plus the Lagrange 
multipliers, constitute a set of parameters to be determined by
minimizing the total energy $\langle H \rangle$. In principle, it is
necessary a set of seven self-consistent parameters.
Once again, as in the infinite $U$ case, the problem was reduced to a one-body
Hamiltonian whose energy can be minimized easily. 
The quantity we need to calculate is the Green function at the impurity, 
around the Fermi level. Thus,
\begin{equation}
G_{dd}^\sigma = \lgr z_\sigma d_\sigma ; d_\sigma^\dagger z_\sigma^\dagger \rgr,
\end{equation}
which is the propagator that carries the correct weight of the Kondo resonance.

In Sec.~\ref{sec:EOM} it will be shown how to calculate the GF  
in the lead's sites using $G_{dd}^\sigma$ as an input.

\subsubsection{Numerical renormalization group approach} 

The NRG method was originally proposed by K. G. Wilson to study magnetic impurity
problems.\cite{Wilson75}  
Initially, it was applied to the Kondo Hamiltonian, and later extended 
to the Anderson model.\cite{Krsihna-murty80} 
It can be shown that for these two models, at low temperatures, the states close 
to the Fermi level (i.e. with the lowest energy contribution) are the most relevant.
Therefore, perturbation theories are not the most adequate approach to these problems.
As a brief description of the method (a full detailed description 
can be found in Refs.~\onlinecite{Wilson75,costi}), we present the two 
main steps in the implementation of the method.\cite{costi} The first one
consists 
in sampling the energy interval of the conduction band by a set of logarithmically 
decreasing energy intervals $[x_N,x_{N-1}]$, defined by $x_N=\pm D\Lambda^{-N}$, 
where $\Lambda$ is the discretization parameter and $D$ is the half-width of the 
conduction band. Then, from each interval, only one representative energy 
value is kept (chosen according to a well defined criterion, see Bulla in Ref.~\onlinecite{costi} 
for details). The total number of representative energies, one from each interval,  
result in the set of discrete energies that couples to the impurity.
After these two basic steps, the total Hamiltonian is mapped into a semi-infinite chain, 
commonly known as {\it Wilson-chain}, 
where each site of the chain corresponds to an energy scale in the logarithmically 
discretized conduction band, with the impurity sitting at its first site. 
It is important to notice that the $t_n$ couplings, between adjacent sites $n$ 
and $n+1$, decrease, away from the impurity, as $\Lambda^{-n/2}$.
The final form for the Hamiltonian in the NRG framework is
\begin{equation}
H = \lim_{N \to \infty} \Lambda^{-(N-1)/2}~H_N  
\label{NRG-Hamil}
\end{equation}
where,

\begin{eqnarray}
H_{N} = \Lambda^{(N-1)/2} \left[ H_{\rm imp} + t' \sum_\sigma \left(
d_\sigma^\dagger c_{0\sigma} + \mbox{h.c.} \right) 
\right. \nonumber \\
+ \left. \sum_{n=0,\sigma}^N \epsilon_n c^\dagger_{n\sigma} c_{n\sigma} +
\sum_{n=0,\sigma}^{N-1} t_n \left( c_{n\sigma}^\dagger
c_{n+1\sigma}+\mbox{h.c.} \right)\right] \label{HNRG}
\end{eqnarray}

where $d_\sigma$ annihilates an electron with spin $\sigma$ at the impurity, 
and $c_{n\sigma}$ annihilates one at site $n$ in the semi-infinite chain 
(indexed from $n=0$ to $N$).

Note that an explicit analytical expression for $t_n$ in Eq.~\ref{HNRG} cannot 
be obtained for a band of arbitrary shape.
For the present problem, where a semi-elliptical band is used, we are forced to 
calculate the $t_n$ numerically.\cite{Chen} 
The hoppings $t_n$ that define the Wilson-chain must not be confused with
the matrix elements $t$ of the real space chain, shown in Fig~\ref{fig1}. The elements
$t_n$ correspond to the band obtained after the logarithmic discretization of the
real space chain. It can be shown that when $\Lambda \to 1$, the hoppings $t_n \to t$.\cite{Chen}

The second important step consists in solving numerically the resulting 
Hamiltonian given by Eq.\ref{NRG-Hamil}. 
To this end, we start with a system consisting of the 
isolated impurity, described by the Hamiltonian $H_{\rm imp}$. Then, 
the subsequent sites are added one by one. This procedure generates 
a sequence of Hamiltonians $H_N$, which are solved as follows: at a given 
iteration $N$ the Hamiltonian $H_N$ is diagonalized numerically. 
The eigenvectors and the corresponding eigenvalues are obtained. Next, a new 
site $N+1$ is added. 
This is done by enlarging the current Hilbert space (associated to iteration $N$) through a tensorial 
product of its elements with the states of the site being added in the next iteration.
This process results in an exponential growth of the dimension of the Hilbert space 
of successive iterations. Due to computational constraints, 
it is necessary to truncate the Hilbert space at each iteration, after it reaches a certain 
size. The NRG truncation criterion is to keep only the $M$ lowest energy states of $H_N$ (typically, $M=1000$), and 
neglect the higher energy spectrum.

The process of adding a single site to $H_N$  
is repeated until the system reaches the strong coupling fixed point. When this fixed point is reached, 
$H_N$ and $H_{N+2}$ have the same eigenvalues.\cite{Wilson75}

The sequence of iterations described above can be thought of as a 
renormalization group (RG) process. 
Adding one site to the chain, and 
obtaining the new low-energy spectra, can be understood as an RG 
transformation ${\cal R}$ that maps the Hamiltonian $H_N$ into a 
new Hamiltonian $H_{N+1}={\cal R}(H_N)$, which has the same form as $H_N$. 
Once the fixed points are obtained, the static and dynamic 
properties, as well as temperature effects, can be 
calculated.\cite{Wilson75,costi} 
In particular we are interested in the local GF at the impurity.

At this point it is worth to remind the 
reader that the information about the high energy dynamics is not 
accurately taken into account, since the high-energy spectra is 
partially neglected after the truncation.

All the NRG data presented in this work was calculated with $\Lambda=2.5$ 
and keeping the $M=1000$ lower energy states in each iteration. To calculate the 
LDOS at the impurity, the delta functions were broadened using Logarithmic-Gaussians 
with a $b=0.6$ factor (see Ref.~\onlinecite{costi}).

Finally, we want to stress, once again, the difference between the real 
space semi-infinite chain and the Wilson-chain. 
The first one, shown in Fig.\ref{fig1}, has all the hopping terms equal to $t$.  
The Wilson-chain is just used to calculate the GF at the impurity, and it is obtained after
the discretization of the real space chain.
With the impurity propagator, obtained from NRG, the LDOS at any site of the real space 
chain can be calculated, as explained below.

\subsection{LDOS away from the impurity: Equation of Motion}
\label{sec:EOM}
In this section, we will explain how to calculate the LDOS at any site of the
semi-chain, which models the electron reservoir.
Note that, for all numerical methods used in this work, once the 
dressed GF is known at the impurity (and, in the case of ECA, for all the other sites of the 
cluster), a procedure based on the construction of a Dyson equation, through 
the use of a sequence of Equation of Motion, can yield the dressed GF 
for any site in the tight-binding semi-chain, no matter how far away from 
the impurity. This can be most easily understood in the case of
ECA, 
as this idea is built into the very core of the method. Indeed, 
ECA allows us to calculate not just the LDOS in all the sites of the cluster, 
but also in all the sites in the
rest of the semi-infinite tight-binding chain used to represent the lead.
An important fact that we want to remark regarding ECA is that the embedding procedure 
results in a {\it feedback} of the leads into the central region, but also reciprocally. 
The physics under study does not need to be restricted to entirely occurring
within the exactly solved region. 
I.e., many-body effects taken into account {\it exactly} inside the ED
cluster 
are propagated, by the Dyson equation, into the electron reservoir (the semi-chain), 
which now does not have anymore the LDOS of a non-interacting system. It is 
important to remark, as will be clearly explained shortly, that the change of the LDOS 
in the semi-chain from tight-binding to many-body lies 
at the core of the method used in this work to estimate the range of the Kondo cloud. 
One added benefit of the procedure to be described below is that the physics 
of the Kondo effect {\it at} the impurity (the Kondo resonance) does not need 
to be calculated with ECA for the EOM procedure to work. In the present work, 
it is also calculated with FUSBMF and NRG.

To calculate the dressed propagators at any site of the semi-chain, 
we write down the EOM of the 
local propagators at a site $M$. In the case of ECA, the site $M$ must be outside 
the cluster, i.e., $M \geq L+1$ (see Fig.~\ref{fig1}). This 
restriction does not apply to FUSBMF or NRG, where the equivalent to the ECA cluster 
can be considered to be {\it just} the impurity.
A brief description of the EOM method can be found in Ref.~\onlinecite{Zubarev}.
To simplify the notation, in what follows we will ignore the spin index $\sigma$. 
The set of equations to solve, in order to calculate $G_{M,M}$, is given by
\begin{eqnarray}
G_{M,M} &=& g_0 + g_0 t~ G_{M-1,M} + g_0 t~ G_{M+1,M} \label{EOM-eq1}
, \\
G_{M+1,M} &=& \tilde{g}_{sc} ~t~ G_{M,M} \label{EOM-eq2}
, \\
G_{M-1,M} &=&  ~G_{M,M-1} \nonumber \\
          &=&  g_{0} t~ G_{M-1,M-1} + {g}_{0} t~ G_{M+1,M-1} \label{EOM-eq3}
\label{gmenos1}
, \\
G_{M+1,M-1} &=& \tilde{g}_{sc} ~t~ G_{M,M-1} , \label{EOM-eq4}
\end{eqnarray}
where $g_0=1/\omega$ is the atomic GF at site $M$ and $\tilde{g}_{sc}$ is the
bare propagator for the rest of the semi-chain starting at the site $M+1$ and is given by,
\begin{equation}
\tilde{g}_{sc} = \frac{\omega \pm \sqrt{\omega^2-4t^2}}{2t^2}.
\end{equation}
In Eq.\ref{gmenos1}, we used explicitly the equivalence between $G_{M-1,M}$ 
and $G_{M,M-1}$. This is only valid if the hopping parameters $t$ are real (e.g., no
magnetic field inside the chain, although a more general EOM, involving a magnetic field, 
can also be found.

Solving this set of equations, we obtain,
\begin{equation}
G_{M,M} = \frac{g_{0} + \displaystyle \frac{g_0 t g_0 t}{1-g_{0}t^2\tilde{g}_{sc}} ~G_{M-1,M-1}}
{1-g_{0}t^2\tilde{g}_{sc}}.
\label{GMM}
\end{equation}
Note that the GF at the site $M$ can be calculated after the GF at $M-1$. 
Note that the equation above 
clearly indicates that, to calculate the dressed GF at site $M$, the only 
many-body information needed is the dressed GF at site $M-1$. 
This fact automatically defines a procedure to find the propagator at any site in the semi-chain.
Note that Eq.~\ref{GMM} is defined for a site, within the semi-chain (i.e. $M \geq 2$), 
that is connected to both adjacent sites by a matrix element $t$. 
Thus, this still leaves us with the task of calculating $G_{1,1}$. 
To calculate the correct propagator at site 1, we have to rewrite 
Eqs.~\ref{EOM-eq1} to~\ref{EOM-eq4}, in order to obtain $G_{1,1}$ as a function of $G_{0,0}$, 
without overlooking that the hopping between sites 0 and 1 is $t'$, not $t$.

For FUSBMF and NRG, we start with the GF calculated at the impurity, i.e., $G_{0,0}$. Using the EOM method, we calculate 
the propagator at the first site of the chain, $G_{1,1}$. Then, using Eq.~\ref{GMM},
the propagator $G_{M,M}$ can be calculated at any site.
The procedure for the ECA method is slightly different, as in ECA all the dressed propagators
inside the ED cluster are calculated already within the method. Therefore, in ECA, 
the EOM procedure starts at site $L+1$ (see Fig.~\ref{fig1}), 
using Eq.~\ref{GMM}, where $G_{L,L}$ is an input from the ECA calculations. 
In that case, there is no special procedure to calculate $G_{1,1}$.
 
Once $G_{M,M}$ is calculated for the desired site $M$, the LDOS can be calculated as,
\begin{equation}
\varrho_{M}(\omega) = \frac{-1}{\pi}~\mbox{Im}[G_{M,M}].
\end{equation}

As we are using the same 
procedure to find the LDOS away from the impurity for three very diverse numerical methods 
(ECA, FUSBMF, and NRG), some explanation about the adopted terminology is necessary, so that the same term, 
with slightly different meanings, can be unambiguously used throughout the manuscript. 
As explained in Fig.~\ref{fig1}, in ECA, {\it cluster} means a variable size finite group 
of sites (including the impurity), which is exactly diagonalized and embedded (as explained above). 
In this manuscript, the ECA {\it cluster} contains up to $L+1=10$ sites 
(i.e., the impurity plus up to $L=9$ tight-binding sites). For sites $N \geq L+1$, the LDOS 
will be found through the EOM method, as described above. On the other hand, an FUSBMF or an NRG 
{\it cluster}, given the very nature of both methods, contains just the impurity itself 
(therefore, $L=0$, see Fig.~\ref{fig1}). Because of that, there is a 
slight difference to the application of the EOM method to these last two 
methods, viz., $G_{1,1}$ has to be calculated first, and then all the other $G_{i,i}$ 
are calculated by using Eq.~\ref{GMM} in sequence, as explained above. 
 
\begin{figure}
\epsfxsize=8.5cm \centerline{\epsfbox{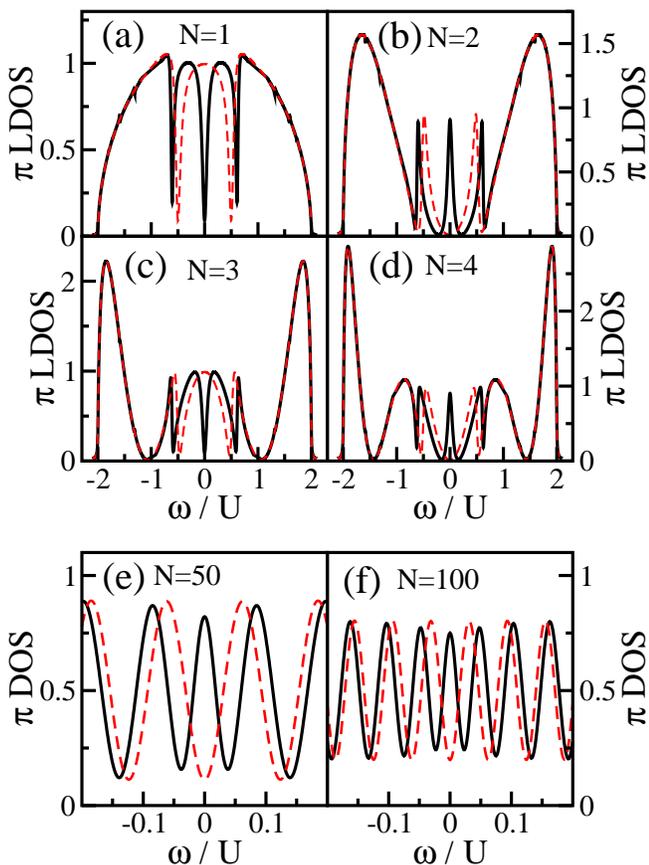}}
\caption{(Color online) (a-f) The LDOS for sites $N=1$,~2,~3,~4,~50, and~100 (see Fig.~\ref{fig1}), 
calculated by ECA for a cluster with $L=3$ (note that the impurity is located at $N=0$). Solid (black) curves show 
the LDOS at $T=0$ for $V_g=-U/2$, $U=t$, and $t'=0.3t$, while dashed (red) curves show 
results for $T>T_{\rm K}$ and the same values for $V_g$, $U$, and $t'$. 
An imaginary part $\nu=0.001$ was used to regularize the LDOS. 
Note that the LDOS's for the first three panels were calculated using ED, in contrast 
to the ones for the last three panels, which were calculated using the EOM method 
described in the text. The LDOS for the first three sites ($N=1$, 2, and 3) was also 
calculated with EOM and, as expected, there was a very good agreement between 
the results obtained with the two different methods.}
\label{fig2}
\end{figure} 

\section{Local density of states within the metal host}
\label{sec:LDOS}

Using the FUSBMF approximation, we can obtain analytically self-
consistent expressions for  the GF, and then the LDOS. The local GF is
also obtainable within NRG, i.e., the LDOS {\it at} 
the impurity can be found with very good accuracy, and, as explained in
Sec.\ref{sec:EOM}, 
the GF (and therefore the LDOS) can be calculated within the semi-chain 
that models the non-interacting band. The same is valid for ECA, despite the distinctions drawn 
above between ECA, in the one hand, and FUSBMF and NRG, in the other hand. 

\begin{figure}
\epsfxsize=8.5cm \centerline{\epsfbox{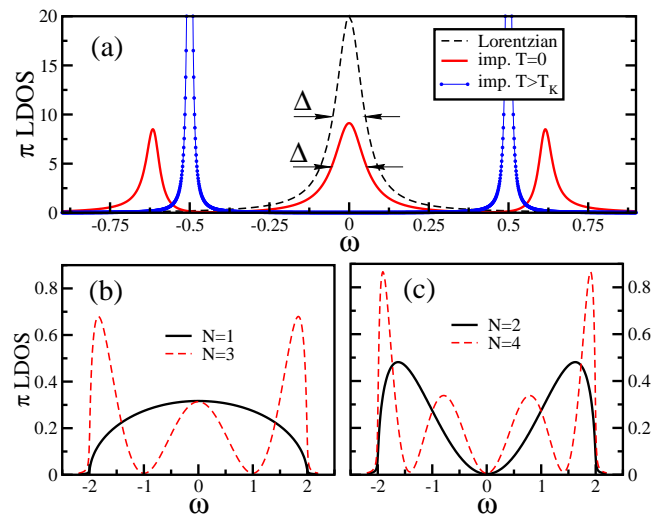}}
\caption{(Color online) (a) LDOS for site $N=0$, where the impurity is located (see Fig.~\ref{fig1}),
calculated by ECA. Solid line shows LDOS for $V_g=-U/2$, $U=t$, and $t'=0.3t$, while in dashed lines,
results are shown for $T>T_{\rm K}$ and the same parameter values for $V_g$, $U$ and $t'$.
An imaginary part $\nu=0.001$ was used to regularize the LDOS. The dashed (black) curve 
displays an example of the normalized Lorentzian used to convolute the data in Eq.~\ref{DOS_distortion}. 
(b) LDOS for the first and third sites of a non-interacting semi-chain
with bandwidth $D=4t$. (c) Same as (b), but now for the second and fourth sites.
}
\label{fig3}
\end{figure}

In Fig.~\ref{fig2}, we show the LDOS for 
several different sites within the non-interacting semi-chain ($N > 0$), 
calculated by ECA. The parameters used are $U=t$ and $\Gamma=0.1t$, and the gate 
potential is set at the electron-hole symmetric point ($V_g=-U/2$).
The LDOS for sites {\it outside} the cluster were obtained through the 
procedure described in Sec.\ref{sec:EOM}. 
The LDOS at zero-temperature is shown in solid lines. 
In dashed lines, for comparison, we show the LDOS when the impurity 
is out of the Kondo regime ($T>T_{\rm K}$). 
In order to obtain a solution for $T>T_{\rm K}$, 
we use the Hubbard-I approximation,\cite{Hubbard} which artificially eliminates the 
spin correlations between the impurity and the leads. This approximation 
is equivalent to performing an ECA calculation where the cluster 
contains just the impurity ($L=0$ in Fig.~\ref{fig1}, i.e., the atomic solution).
To understand the results in Fig.~\ref{fig2}, it is instructive to
analyze the 
results shown in Fig.~\ref{fig3}, where, in panel (a) it is shown the LDOS 
at the impurity, for $T=0$ [(red) solid line) and for $T > T_{\rm K}$ [(blue) dotted curve], 
and in panels (b) and (c) the 
LDOS of the first four edge sites of an isolated non-interacting semi-chain. 
The results shown in Fig.~\ref{fig2} [panels (a) to (d)] display essentially the {\it hybridization} 
between the LDOS of the impurity (panel (a) in Fig.~\ref{fig3}) 
and the LDOS of the sites in the semi-chain (panels (b) and (c) in Fig.~\ref{fig3}), 
once the impurity is coupled to the semi-chain. 
The solid (black) curve shows the LDOS for $T=0$, while the dashed (red) curve shows the 
LDOS for $T > T_{\rm K}$, at the first four sites in the semi-chain, after it couples to the 
impurity. 
This hybridization can be described in a simple way: a peak in the LDOS of the impurity [Fig.~\ref{fig3}(a)], 
centered at $\omega_p$, will generate either a resonance or an anti-resonance (at $\omega_p$) 
in the LDOS of a semi-chain site when the impurity couples (hybridizes) to the semi-chain. 
On the one hand, a resonance (a peak) will result if the semi-chain's LDOS, in one specific site, 
vanishes at $\omega_p$. On the other hand, an anti-resonance (a dip) results when the site's LDOS at $\omega_p$ is finite. 
This resonance/anti-resonance site to site oscillation effect in the LDOS 
will have important consequences in the next section.
Our interest is to be able to distinguish the effect caused over the semi-chain's LDOS, far away from the impurity, by the 
presence ($T < T_{\rm K}$) or absence ($T > T_{\rm K}$) of a resonance at the Fermi energy 
(the Kondo peak) in the LDOS {\it at} the impurity [compare the solid and dotted curves in Fig.~\ref{fig3}(a)]. 
The extent to which this hybridization effect can spread away from the impurity will 
be used as a measure of the extent of the Kondo cloud. 

\begin{figure}
\epsfxsize=8cm \centerline{\epsfbox{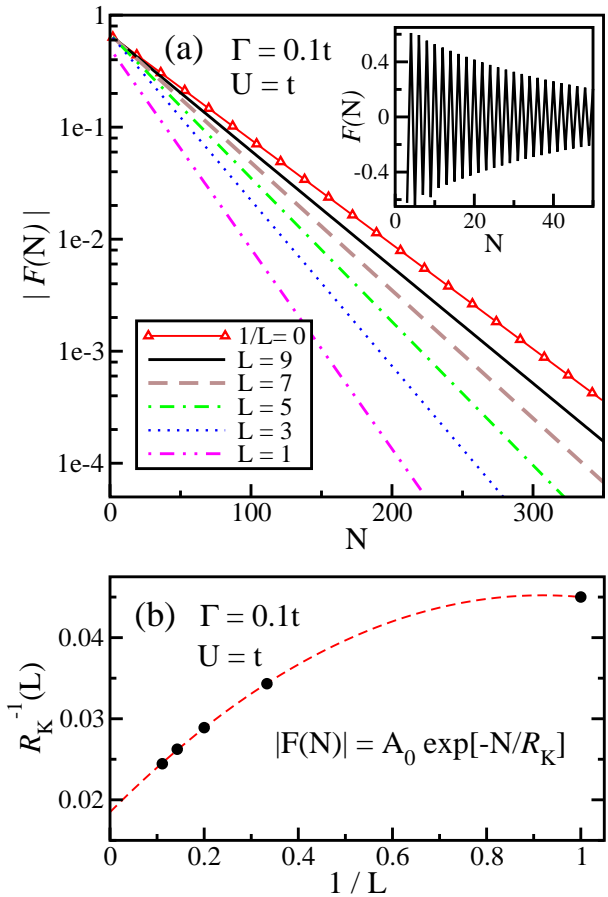}}
\caption{(Color online) (a) Main panel: Absolute value of the cloud extension function, $|F({\rm N})|$,
calculated by ECA as a function of the semi-chain site $N$ for several different cluster
sizes ($L=1$ to 9, in steps of 2 -- see legend). The parameter values are $U=t$ and $\Gamma=0.1t$. Note that
$L$ is the number of nearest neighboring sites to the impurity within the cluster solved exactly in ECA.
The curves shown are the data obtained through Eq.~\ref{DOS_distortion}. 
Fits to these data (not shown), for each value of $L$, using Eq.~\ref{exponential}, result in exactly the same curves 
as the ones shown. 
It is clear that the data for each different cluster size decays exponentially with N (notice the
logarithmic scale in the vertical axis). The open (red) triangles curve shows the extrapolation
of $|F({\rm N})|$ to the thermodynamical limit, as describe in detail in the text.
The inset shows the site to site oscillations of $F({\rm N})$, as well as the
exponential decay, now with a linear scale for the vertical axis.
(b) ECA  extrapolation to the thermodynamical limit (dashed (red) curve), as described in the text,
of the $R_K$ data (solid (black) dots) obtained from the fittings for each different $L$ curve
shown in panel (a), as explained in the text.}
\label{fig4}
\end{figure}

Figure~\ref{fig2} clearly displays this kind of hybridization effect, as described above. Indeed, 
if one concentrates the attention on the features close to the Fermi energy ($\omega=0$) 
in the different panels in Fig.~\ref{fig2}, one sees that the difference between the LDOS curves below $T_{\rm K}$ 
(solid) and above $T_{\rm K}$ (dashed) is quite marked, and owes its origin to 
the presence of the Kondo peak at the impurity below $T_{\rm K}$. By using the Lorentzian shown in 
Fig.~\ref{fig3} (dashed line) to restrict ones attention to the immediate neighborhood 
of the Fermi energy, by convoluting it with the {\it difference} between the solid and 
dashed curves in Fig.~\ref{fig2}, one expects to extract the essence of the influence of the impurity, 
when in the Kondo regime, over the Fermi sea. One can picture the change from the 
solid to the dashed curve, say, in site 50 [panel (e)], as that occurring in the LDOS away from the 
impurity when the temperature is lowered below $T_{\rm K}$. Panel (f), where there is very little 
difference between both curves, shows that the impurity, in a Kondo regime, has a 
spatially limited influence over the Fermi sea. It is one of the aims of this paper to understand how 
this influence depends on the sole energy scale of the Kondo effect, i.e., the Kondo 
temperature $T_{\rm K}$.

\section{Numerical results}
\label{sec:results}

\begin{figure}
\epsfxsize=8cm \centerline{\epsfbox{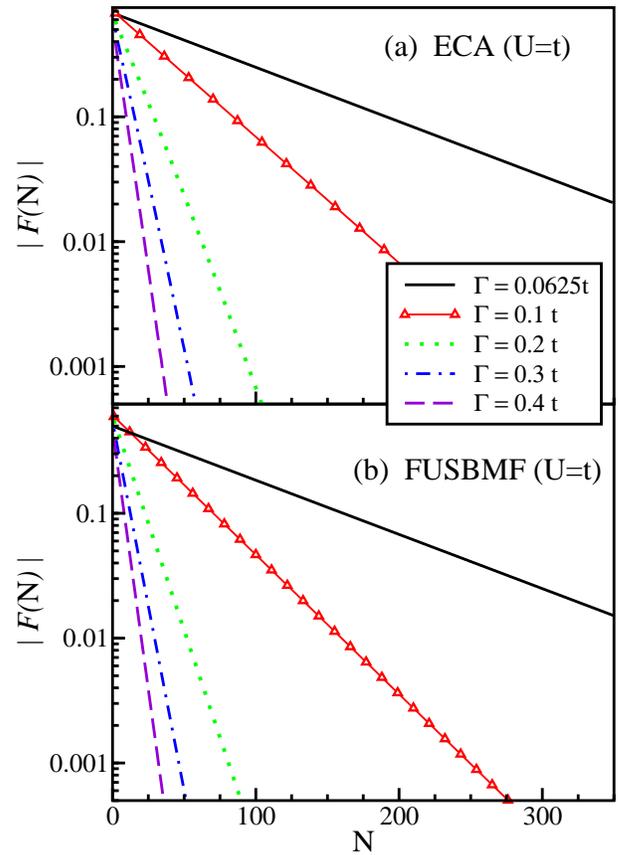}}
\caption{(Color online) Cloud extension function $F(N)$ as a function of the site in the semi-chain, 
for several values of $\Gamma$. (a) Extrapolation to the thermodynamical 
limit  for each $\Gamma$ value and (b) FUSBMF results.
Note that, as $\Gamma$ increases, the reach of the distortion in the LDOS of a given site $N$ ($R_{\rm K}$), 
produced by the impurity, is reduced, reflecting the shorter range of the Kondo effect. 
Note also that the dependence of $|F(N)|$ with $N$ for all $\Gamma$ values shown is perfectly linear.}
\label{fig5}
\end{figure}

\begin{figure}
\epsfxsize=8cm \centerline{\epsfbox{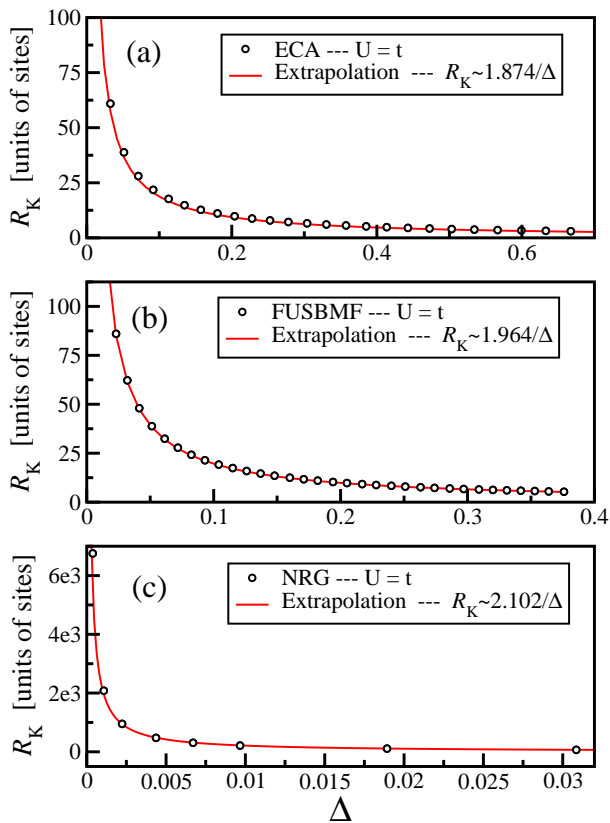}}
\caption{(Color online) Range ($R_{\rm K}$) of the Kondo cloud as a 
function of $\Delta$. The open dots show the results obtained for $R_{\rm K}$ through the fitting 
of curves like the ones in Fig.~\ref{fig5} for various values of $\Gamma$ (therefore, different $\Delta$, 
i.e., (the width of the resonance at the Fermi level $\Delta$). 
(a) ECA, (b) FUSBMF,  and (c) NRG results.
A fixed parameter $U=t$ was used for the three methods, and the parameter $\Gamma$
was changed in order to obtain different $\Delta$ values.  
The solid (red) line in each panel shows the interpolation of an $1/\Delta$ function, as 
expected for $R_{\rm K}$ vs $T_{\rm K}$ (which is proportional to $\Delta$). 
The value of the proportionality factor for each method is shown in the respective label.}
\label{fig6}
\end{figure}

\begin{figure}
\vspace{0.05cm}
\epsfxsize=8.5cm \centerline{\epsfbox{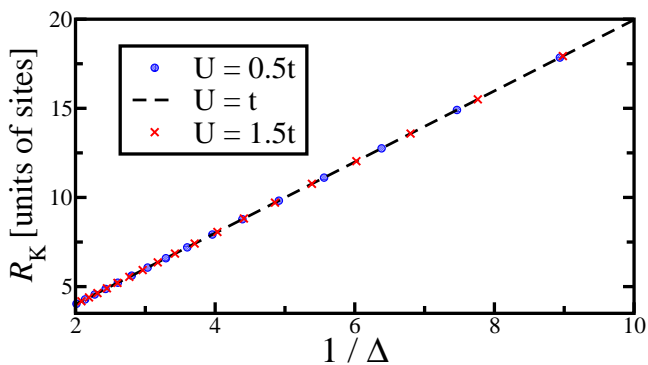}}
\caption{(Color online) ECA results for the size of the Kondo cloud vs. $1/\Delta$ for $U=0.5t$, $U=t$ and $U=1.5t$.  
Note that, for the interval of values of $U$ and $\Gamma$ shown, $R_{\rm K}$ is exactly a linear 
function of $1/\Delta$, and does not depend on the values of $U$ and $\Gamma$ used.}
\label{fig7}
\end{figure}

As mentioned in Sec.~\ref{sec:introduction}, considering that the width of the Kondo
resonance in the LDOS at the impurity, $\Delta$, is proportional to $T_{\rm K}$,
we expect that,
\begin{equation}
R_{\rm K} \approx \frac{1}{\Delta},
\label{R_K2}
\end{equation}
In this section, we will estimate the screening length $R_{\rm K}$ 
by evaluating the {\it distortion} in the LDOS, produced by the Kondo resonance at the impurity, 
in sites N arbitrarily far away from the impurity. The distortion produce in site N will be 
quantified by the absolute value of the function $F(N)$, as defined by 
\begin{equation}
F({\rm N}) = \int_{-\infty}^{\infty} \left( \varrho_{\rm N}^{\rm K}(\omega)
 - \varrho_{\rm N}^{\rm{NK}}(\omega) \right) ~L_{\Delta}(\omega)~d\omega ,
\label{DOS_distortion}
\end{equation}
where $\varrho_{\rm N}^{\rm K}$ ($\varrho_{\rm N}^{\rm{NK}}$)
is the local density of states at the site N in the Kondo regime
(out of Kondo), and $L_{\Delta}$ is the Lorentzian distribution of width $\Delta$ 
[dashed curve in Fig.~\ref{fig3}(a)]. Note that the distance to the impurity is 
given by $r=Na$, where $a$ is the lattice parameter. It is important at this point to 
remind the reader that all the calculations here were done at $V_g=-U/2$ (i.e., at the 
particle-hole symmetric point). 
To evaluate $\varrho_{\rm N}^{\rm{NK}}$, we applied the EOM procedure to the Hubbard-I solution 
for the impurity's GF (where the dotted line in Fig.~\ref{fig3}(a) shows the negative of its imaginary part).

Note that, for a site far away from the impurity, where the many-body effects are
not important, $\varrho_{\rm N}^{\rm K}$ must be equal to
$\varrho_{\rm N}^{\rm{NK}}$, thus $F({\rm N}) \approx 0$. This can be seen in the 
last panel of Fig.~\ref{fig2}. $F({\rm N})$ will be used to find a length scale
beyond which the presence of the impurity is not relevant any more. We will call
$F({\rm N})$ the {\it cloud extension function}.

As already mentioned, to calculate $\varrho_N^{\rm K}$, we will use ECA, FUSBMF, and NRG. 

In Fig.~\ref{fig4}(a), it is shown the absolute value of the cloud extension function, $|F({\rm N})|$, 
calculated with ECA, as a function of $N$, for $U=t$, $\Gamma=0.1t$,
and for several values of $L$ (note that the vertical axis has a 
logarithmic scale). The scatter plots
show the data obtained from Eq.~\ref{DOS_distortion} for some selected values of N, for different
cluster sizes $L$ (see legend).
We find that the behavior of $|F({\rm N})|$, for all values of $L$ used,
is a decaying exponential (the correlation factor, when
fitting the curves with an exponential, was exactly 1 for all values of $L$): 
\begin{equation}
|F({\rm N})| = A_0\exp(-N/R_{\rm K}).
\label{exponential}
\end{equation}
where $R_{\rm K}$ marks the distance from the impurity where the value of $|F({\rm N})|$ has
fallen by $1/e$, in comparison to its value at the impurity ($N=0$).
The solid lines in panel (a) show the result of fitting each set of data with 
Eq.~\ref{exponential}. 
In the inset of Fig.~\ref{fig4}(a), it can be observed
that $F({\rm N})$ oscillates between positive and negative values for successive N.\cite{note3}
This is a direct consequence of the resonance/anti-resonance oscillation 
in the LDOS discussed in the previous section. 
The extrapolation of $R_{\rm K}$ to an `infinite' ECA cluster ($1/L \to 0$) is explained in detail next.

As mentioned already, each curve in Fig.~\ref{fig4}(a) (from $L=1$ to $L=9$) is fitted using Eq.~\ref{exponential}.
Therefore, a value for $R_{\rm K}(L)$ is found for each cluster size used in ECA ($1 \leq L \leq 9$). 
Figure~\ref{fig4}(b) shows the values obtained this way for $R_{\rm K}(L)$ as a function of $1/L$ (solid (black) dots), 
for the different values of $L$ used in panel (a). A fitting of these results 
by a quadratic polynomial is also shown (dashed (red) curve). The intercept of the dashed (red) 
curve with the vertical axis provides an extrapolation of $R_{\rm K}$ to the thermodynamical limit 
$R_{\rm K}(L \to \infty)$.
This extrapolated value of $R_{\rm K}$ can then be used in Eq.~\ref{exponential} to obtain 
the thermodynamical limit for the $|F(N)|$ curve, which, for $\Gamma/t=0.1$, is the 
open (red) triangles in Fig.~\ref{fig4}(a).\cite{extrapolation2} 

Obviously, Eq.~\ref{exponential} 
has two free parameters, viz., $A_0$ and $R_{\rm K}$. Although the vertical axis in 
Fig.~\ref{fig4}(a) is logarithmic, making it difficult to judge the 
convergence of $A_0$, it is true that $A_0$ converges with $L$ faster than 
$R_{\rm K}$ (note that, in accordance with Eq.~\ref{exponential}, $A_0$ is 
the y-intercept and $R_{\rm K}$ is the negative of the inverse of the slope of the curves for different 
cluster sizes). The values of $A_0$ for different values of $L$ where 
obtained from the fitting, of each data set in Fig.~\ref{fig4}(a), done with Eq.~\ref{exponential}.  
The dependence of $A_0$ on the model parameters will be discussed below.\cite{note4}

At this point, it is important to note that, as the cluster used in the FUSBMF and NRG calculations 
has a fixed size ($L=0$), there is no extrapolation to be done to find 
$|F(N)|$ for both methods. There is only a fitting to Eq.~\ref{exponential} 
to find $R_{\rm K}$ and $A_0$. 

Now that we have clarified how the thermodynamical limit value for $R_{\rm K}$ is found 
for each method, we want to show how it varies with the model parameters. 
Figure~\ref{fig5}(a) shows the extrapolated (ECA) $|F(N)|$ curves, for $U=1.0$ and 
different values of $\Gamma$ (from 0.0625 (solid (black) curve) to 0.4 (dashed (magenta) curve). 
From the data, it is clear that $R_{\rm K}$ (the negative of the inverse of the slope) 
decreases with increasing values of $\Gamma$. 
In panel (b), the corresponding $|F(N)|$ 
curves obtained with FUSBMF are shown, for comparison. The overall agreement between 
both methods is quite good. As expected, we observe that the size of the Kondo 
cloud (measured through the cloud extension function) increases with $U/\Gamma$. 
We can understand this behavior by noting that, as $\Gamma$ increases (with a fixed $U$), 
$T_{\rm K}$ also increases, and $R_{\rm K}$, as predicted by Eq.~\ref{R_K}, decays.

Figure~\ref{fig6} shows the results obtained [open dots in panels (a), (b) and (c)] by 
extrapolating $R_{\rm K}$ from $|F(N)|$ for different values of $\Gamma$ (at $V_g=-U/2$), 
using ECA (a), FUSBMF (b), and NRG (c). 
These results are plotted as a function of $\Delta$ (which is taken as the full-width at half-height 
of the Kondo peak for each different value of $\Gamma$. 
We observe that the dependence of the Kondo length $R_{\rm K}$ with $\Delta$ satisfies the 
relationship given by Eq.~\ref{R_K2} (as $\Delta$, the width of the Kondo peak, is proportional to $T_{\rm K}$). 
To emphasize that, each set of data (obtained by the three different methods) was fitted by a function $\propto 1/\Delta$ 
(see the solid (red) line in each panel). It is important to stress that the proportionality coefficient 
between $R_{\rm K}$ and $1/\Delta$ obtained by all the three different methods is very similar, 
i.e., $R_{\rm K} \sim 2.0/\Delta$. The proportionallity factor in Eq.~\ref{R_K2} are
$1.874$ for ECA, $1.964$ for FUSBMF, and $2.102$ for NRG.
While this factor is similar for ECA and FUSBMF, there is a 10\% difference between ECA 
and NRG. We believe that this difference comes from the parameter $b$ used in NRG 
to broaden the logarithmic-Gaussian functions in the LDOS, as the value of $\Delta$ 
obtained by NRG is very sensitive to the choice of this arbitrary parameter.

{Figure~\ref{fig7} shows ECA results of $R_{\rm K}$ v.s. $1/\Delta$ 
for three different values of $U$ (0.5, 1.0, and 1.5). 
As shown in the figure, the functional form $R_{\rm K} \propto 1/\Delta$ is valid for the 
intervals of $U$ and $\Delta$ (and therefore $\Gamma$) inside which the calculations were done. 
More importantly, as all the curves collapse to a single line, the proportionality coefficient is 
also independent of these intervals. 
This indicate that the `propagation' of the Kondo effect into the
leads 
(which is essentially measured by $R_{\rm K}$) depends only on the weight of the 
Kondo resonance at the Fermi level (measured by $\Delta$).

Figure~\ref{fig8} shows, for ECA and 
FUSBMF, the parameter $A_0$ as a function of $\Delta$. We can see that, the ECA and FUSBMF
curves agree quite well, for the same value $U=1.0$. Additional calculations, 
with different $U$ values ($U=0.5$ and $U=1.5$), were done just 
with ECA. For these additional results, one sees that the curves start to differ from each other for large
$\Delta$, but agree for small values ($\Delta < 0.04$). The agreement for small $\Delta$ can be easily understood 
if one takes in account the universal behavior of the Kondo effect, in the sense that it 
is determined by a single energy scale, the Kondo temperature $T_{\rm K}$. 
Therefore, $A_0$ (as $R_{\rm K}$) does not depend on either $U$ or $\Gamma$
independently, 
but on their ratio ($U/\Gamma$), at least until the system enters the mixed valence regime, 
at higher $\Delta$ (equivalent to $\Gamma$). Notice that the curve for lower $U$ ($U=0.5$) starts 
to diverge from the other two at a lower value of $\Delta$ (proportional to $\Gamma$), while the 
opposite occurs for the larger-$U$ curve.\cite{note3}

\begin{figure}
\vspace{0.5cm}
\epsfxsize=8cm \centerline{\epsfbox{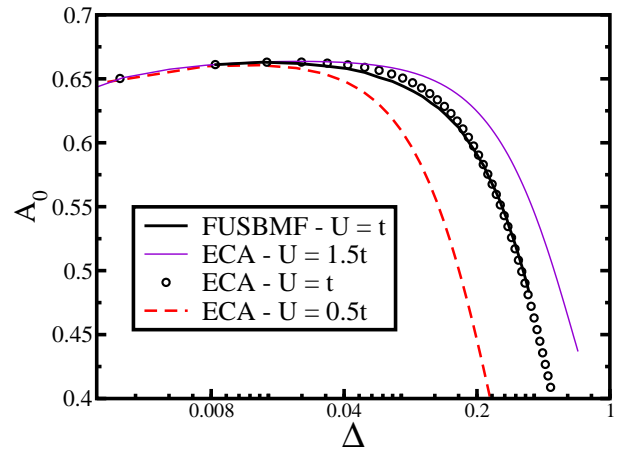}}
\caption{(Color online) Parameter $A_0$ as a function of $\Delta$ (in log-scale) calculated by ECA 
and FUSBMF. Different $U$ where used for ECA. Note that, when $\Delta$ is small, all 
curves coincide, as expected, in view of the universal behavior characteristic of a Kondo system. 
Increasing $\Delta$, we enter in the mixed valence regime, and curves for different values of $U$ 
start to diverge.}
\label{fig8}
\end{figure}

\section{Conclusions}

Using, for the first time, the LDOS within the metal lead, we have estimated 
the effective length, in real space, of the 
effect of the many-body correlations originating at the impurity site. 
For the Kondo effect, we defined a cloud extension function $F({\rm N})$ in order to estimate 
when an electron located at a site $N$, away from the impurity, 
is not affected anymore by its presence.

Indeed, we have used the electronic properties reflected in the LDOS function ({\it charge spectra}) 
of the one-dimensional metallic lead,  
to study the spatial propagation of the Kondo effect away from the magnetic impurity. 
The length of the Kondo cloud, $R_{\rm K}$, has been defined in the 
literature to be the extension of the spin-screening-cloud, formed by the conduction electrons, 
in the vicinity of the impurity. From this point of view, it is essentially the spatial 
size of a {\it magnetic} property, as it is associated to the spin-spin correlations between 
the local impurity and the conduction electron spins.\cite{fabian} 
However,
here we claim that, as far as the Kondo cloud is concerned, the {\it charge spectra} 
counterpart of the Kondo physics (defined as the effect of the impurity's Kondo peak
over the LDOS of the leads) is equivalent to its magnetic expression, as they are both
manifestations of the same physical phenomenon. Moreover, we claim
that $|F(N)|$, being dependent just on the LDOS, is easier to
calculate and measure than spin-correlation-based functions.

As far as transport properties are concerned, if the Kondo effect is 
thought to be the way in which two or more QDs can interfere, the relevant way of 
studying the Kondo cloud is by analyzing the effect of the impurity over the LDOS of the 
rest of the system. 

In order to study the Kondo effect spatial propagation, we define what we call a cloud extension 
function, denoted  $F(N)$ (see Eq.~\ref{DOS_distortion}). It measures, in an interval of 
width $T_K$ around the Fermi energy,  
the distortion of the LDOS, at site N, created by a Kondo impurity sitting at the origin. 
We evaluated this function using three totally different formalisms, 
ECA, FUSBMF, and NRG, and obtained almost identical results for the 
variation of $R_K$ with the parameters of the system. The fact that 
three different formalisms provide the same physical description makes 
this study quite robust and reliable. We demonstrate, as well, that the 
length of the Kondo cloud is controlled 
by the unique, scaling invariant, relevant parameter of the Kondo effect, the 
Kondo temperature $T_K$. These results permit a very accurate determination 
of the functional form of $R_K(T_K)$, in agreement with intuitive ideas, 
summarized in Eq.~\ref{R_K}.

Finally, it is important to emphasize that the measurement of spin-spin correlations 
between different sites, in a real STM experiment, is difficult to 
perform, as the use of two different STM tips, at the same time, is required.\cite{Markus} On the 
other hand, the mapping of the Kondo cloud through the difference in the conductance, measured 
by an STM tip, at different points in a 1D system, looks more feasible, as it has 
already been performed in metallic surfaces.\cite{crommie}

\vspace{0.5cm}
\begin{acknowledgements}
The authors wish to acknowledge fruitful discussions with K. A. Al-Hassanieh, G. Chiappe, 
E. H. Kim, and especially F. Heidrich-Meisner. 
E.V.A. thanks the Brazilian agencies FAPERJ, CNPq (CIAM project), and CAPES for financial support.  
G.B.M. and C.A.B. acknowledge support by NSF Grant No. DMR-0710529. 
E.D. is supported by the NSF Grant No. DMR-0706020 and the Division of Materials 
Science and Engineering, U.S. DOE, under contract with UT-Battelle, LLC. E.V. acknowledges support of CNPq (CIAM project).
\end{acknowledgements}



\begin{thebibliography}{99}
\bibitem{note1} Although one should be cautious in taking this approach, as a Heavy Fermion system 
is modeled by a Kondo Lattice Hamiltonian, while all the results presented here are for a single 
impurity.
\bibitem{note2} Note that, in ECA, $\Delta$ was calculated as the thermodynamical limit
-$1/L \to 0$- of the width of the Kondo resonance at the impurity's LDOS.
\bibitem{note4} The limit in which $|F(N)| \to 0$ determines the order of magnitude of the Kondo cloud.
However, the actual values and functional variation of both $A_0$ and $R_{\rm K}$ in
Eq.~\ref{exponential} clearly indicate that $R_{\rm K}$ has a leading contribution in
determining the size of the Kondo cloud, given the small variation of $A_0$ in the
region of parameter space associated to the Kondo effect (compare the scales of
the vertical axis in Figs.~6 and 8).
\bibitem{note3} Note that $A_0$ was {\bf not} calculated as $|F(0)|$, but through the fittings discussed 
in Figs.~4 and 5. Therefore, its value, and its dependence with $\Delta$, as seen in Fig.~8, do not 
represent the value or the variation of the area of the Kondo peak.
\end{thebibliography}
\end{document}